\documentclass[lettersize,journal]{IEEEtran}
\usepackage{xcolor}
\usepackage{hyperref}
\usepackage{amsmath,amsfonts}
\usepackage{algorithmic}
\usepackage{algorithm}
\usepackage{array}
\usepackage[caption=false,font=normalsize,labelfont=sf,textfont=sf]{subfig}
\usepackage{textcomp}
\usepackage{stfloats}
\usepackage{url}
\usepackage{verbatim}
\usepackage{multirow}
\usepackage{graphicx}
\usepackage{float}
\usepackage{cite}
\begin{document}
	
	
	\title{An All-Optical General-Purpose CPU and Optical Computer Architecture}
	
	\author{Michael Kissner, Leonardo Del Bino, Felix Päsler, Peter Caruana, George Ghalanos
		\thanks{The authors are with Akhetonics GmbH, Akazienstr. 3a, 10823 Berlin, Germany (e-mail: \href{michael@akhetonics.com}{michael@akhetonics.com})}
	}





\maketitle

\begin{abstract}
	Energy efficiency of electronic digital processors is primarily limited by the energy consumption of electronic communication and interconnects. The industry is almost unanimously pushing towards replacing both long-haul, as well as local chip interconnects, using optics to drastically increase efficiency. In this paper, we explore what comes after the successful migration to optical interconnects, as with this inefficiency solved, the main source of energy consumption will be electronic digital computing, memory and electro-optical conversion. Our approach attempts to address all these issues by introducing efficient all-optical digital computing and memory, which in turn eliminates the need for electro-optical conversions. Here, we demonstrate for the first time a scheme to enable general-purpose digital data processing in an integrated form and present our photonic integrated circuit (PIC) implementation. For this demonstration we implemented a URISC architecture capable of running any classical piece of software all-optically and present a comprehensive architectural framework for all-optical computing to go beyond.
\end{abstract}


\section{Introduction}

Optical computers have often been seen as the next step in the future of computing ever since the 1960s\cite{Ambs:2010}. They promise much higher energy efficiencies, while offering near latency free, high-performance computing (HPC), as well as easily scalable data bandwidth and parallelism. Ever since 1957 with von Neumann publishing the first related patent\cite{Patent:1957} and Bell Labs creating the first optical digital computing circuits\cite{Huang:1984,Brenner:1986,Murdocca:1988,Brenner:1988}, the field has gone through multiple hype cycles and downturns\cite{Ambs:2010}. While an all-optical, general-purpose computer hadn't been achieved yet so far, the DOC-II\cite{Guilfoyle:1991} did come close,  still requiring an electronic processing step each iteration. A lot of results from the research since has found its way into use on the optical communication side, with optical communication to enhance electronic computing proving itself to be the most useful and successful development.

There is a very sensible push in industry and research, that the focus should not be on the computation itself, but rather on the communication between compute elements\cite{Miller:2016}. Currently, the energy consumption of electronic communication happening on and off chip can reach over $80\%$ of the total power consumption of an electronic processor\cite{Ho:2023,Horowitz:2014,Das:2015}. By replacing these electronic wires with optical interconnects, one can expect an immediate efficiency gain of up to a factor of $6$, due to waveguides being near lossless and highly efficient electro-optical modulation schemes. As an example, communicating a bit across an electronic chip can cost up to 600 fJ, whereas using electro-optical modulation would enable far lower energies, down to the 1-10 fJ domain\cite{Zhang:2023}, with the potential of reaching the attojoules in the coming years. The industry expects the majority of efficiency gains in the near future coming from such electro-optical hybrid computing approach, with both large companies, like Intel and Nvidia, as well as startups, like Lightmatter, Ayar Labs, Celestial AI and Black Semiconductor, actively developing schemes for optical interposers to connect electronic computing elements and memory. This development is highly beneficial not just for the field of computing, but also photonics in general. 

\begin{figure}[!t]
	\centering
	\includegraphics[width=3.4in]{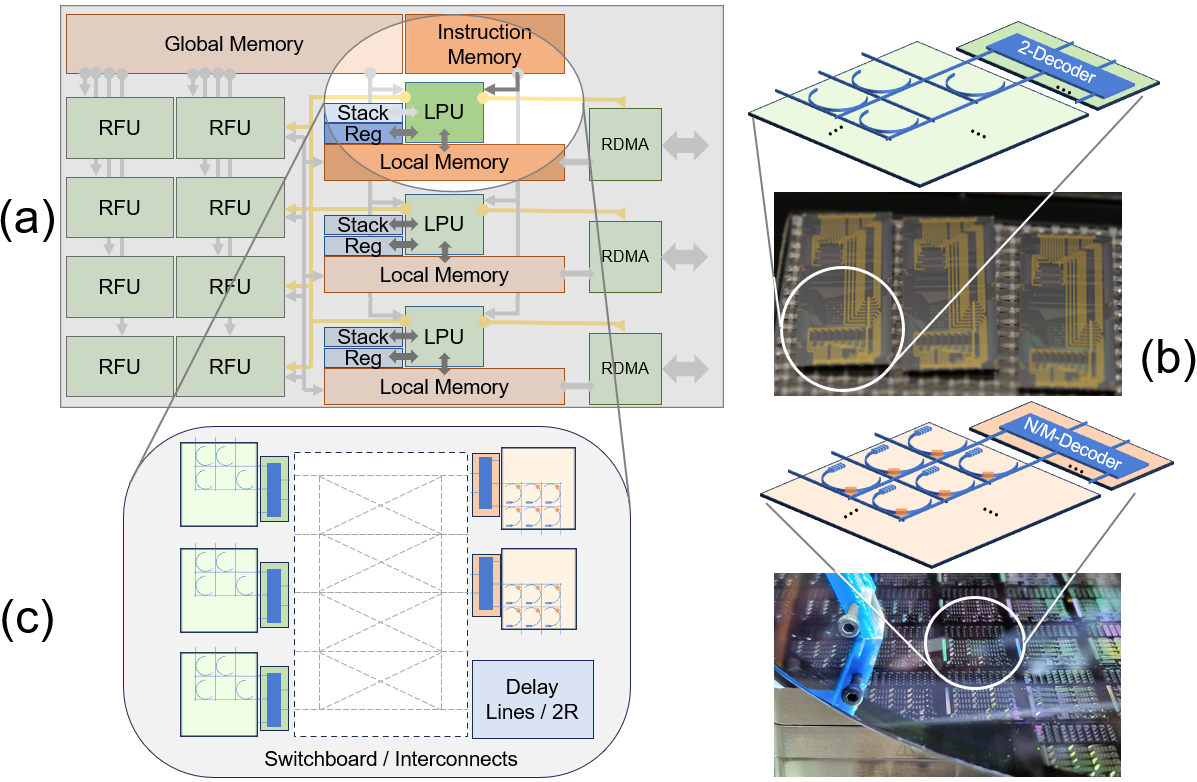}
	\caption{Our all-optical, cross-domain computing architecture in the form of an XPU (a). The core of this architecture is implemented using photonic integrated circuits (PIC) shown in (b), with the focus being on the logic processing unit (LPU) and memories (c), where for this demonstration, we show for the first time how it operates as a stand-alone all-optical CPU.}
	\label{fig:summary}
\end{figure}

With our research, however, we are focused on the phase thereafter. Once optical interconnects and interposers have been fully established and are the main mode of inter-chip and intra-chip communication, solving the $6\times$ inefficiencies, the bottleneck will shift once again. By essentially eliminating the inefficiencies from communication itself, electronic computation and memory will become the main source of power consumption. Using \cite{Miller:2016} as an example again, reducing the interconnect power consumption from 600 fJ to 1-10 fJ per bit, would then result in the compute (100 fJ for a floating point operation in this example) to dominate the energy equation. Our focus should be on reducing the energy consumption of compute and memory either after or in parallel to reducing interconnect power consumption. Through such a combined, all-optical approach, the potential to unlock another $10\times$ efficiency increase in the long term.

Here we concern ourselves with all-optical computing (Figure \ref{fig:summary}), i.e. the data computed on always remains optical and no electro-optical conversion is occurring throughout the device. We do acknowledge that there will of course be electronics in the device for powering lasers or tuning purposes, but they do not participate in the actual processing itself. This is a rather radical departure from the classical electronic or electro-optical hybrid approach, performing some of the arithmetic optically, but control flow and most non-linear arithmetic purely in electronics\cite{McMahon:2023, Tsakyridis:2024}. In such hybrids, as well as electronics with optical interposers, the main R\&D focus lies on the performance and efficiency of interfacing electronics with optics. With an all-optical approach, the focus lies solely on the optics. 

This all-optical form of computing, as we intend to lay out in this paper, has unique advantages over any existing or upcoming electronic schemes, requiring a very different view on how we think about the fundamental building blocks, as well as the computer architectures. Thus, our goal is threefold: 

\begin{itemize}
	\item to present an architectural guide for optical computing beyond von Neumann and the scalability of the approach to surpass what is possible in electronics (Section \ref{sec:arch}),
	\item to clear up common and long held misconceptions on optical computing (Section \ref{sec:arch}), and
	\item to present our implementation of our all-optical, general-purpose 2-bit demonstrator CPU and the feasibility of performing general-purpose digital computing all-optically (Section \ref{sec:impl}).  
\end{itemize}

The paper introduces the required computer architecture concepts as needed and does not assume prior knowledge in the field. Our main contribution will be a novel architecture using a blend of old and new concepts (such as our RWORM, delay line based registers and all-optical LUT3 in section \ref{sec:impl}) for all-optical computing on a photonic platform, that enables feasible general-purpose operation.

\subsection{Digital Optical Computing Past and Present}\label{sec:sota}

Digital optical computing up to the 2000s was focused on using discrete optics to achieve binary logic\cite{Athale:1990}. But with the sheer number of elements needed, it was often seen as too hard to accomplish\cite{Caulfield:1998}. Nonetheless, a lot of important results arose from this era of optical computing, such as Miller's criteria for practical optical logic (Table \ref{tab:miller}). Now, integrated photonics and photonic integrated circuits (PIC)\cite{siew2021review, Shekhar:2024} offer a new path to overcome these challenges and it is vital to have a renewed look at the results from past.

\begin{table}
	\begin{center}
		\caption{Miller's criteria for practical optical logic reproduced\\ from \cite{Miller:2010} (shortened).}
		\label{tab:miller}
		\renewcommand{\arraystretch}{1.2}
		\begin{tabular}{ p{1.7cm} | p{5.4cm} | c }
			Criteria & Description &  \\
			\hline
			Cascadeability & The output of one stage must be in the correct form to drive the input of the next stage. & \parbox[t]{1mm}{\multirow{4}{*}{\rotatebox[origin=c]{-90}{\;\;\;\;essential}}}\\
			\cline{1-2}
			Fan-out & The output of one stage must be sufficient to drive the inputs of at least two subsequent stages (fan-out or signal gain of at least two). & \\
			\cline{1-2}
			Logic-level restoration & The quality of the logic signal is restored so that degradations in signal quality do not propagate through the system; that is, the signal is ‘cleaned up’ at each stage. & \\
			\cline{1-2}
			Input/output isolation & We do not want signals reflected back into the output to behave as if they were input signals, as this makes system design very difficult. & \\
			\hline
			Absence of critical biasing & We do not want to have to set the operating point of each device to a high level of precision. & \parbox[t]{1mm}{\multirow{2}{*}{\rotatebox[origin=c]{-90}{\;\;\;\;optional}}}\\
			\cline{1-2}
			Logic level independent of loss & The logic level represented in a signal should not depend on transmission loss, as this loss can vary for different paths in a system.  & \\
			\hline			 
		\end{tabular}
	\end{center}
\end{table}

An important topic was, of course, the architecture an optical computer would take. Should the architecture be a carbon copy of electronics, such as a von Neumann, or something new? In the late 80's and early 90's a lot of architectures were proposed\cite{Huang:1984, Brenner:1986, Brenner:1988, Cathey:1989, Athale:1990, Jackson:1991, Jackson:1994}, with some being realized, such as the previously mentioned DOC-II and Bell Labs processor. During this time, Sawchuk and Strand explored the use of systolic arrays and crossbars for optical computing\cite{Sawchuk:1984}. In more recent years, optical pushdown automata and optical finite-state-machine, that implicitly try to address one of the core issues of any optical digital computer, namely memory, have been explored\cite{Touch:2016}. We highlight these concepts, as they play an important role in the formulation of our architecture. 

The almost canonical way of performing all-optical switching and logic is to use semiconductor optical amplifiers (SOA) and exploit their cross-gain modulation (XGM) or cross-phase modulation (XPM) capabilities\cite{Toliver:2000}. With very reliable devices having been shown over the past 20 years\cite{Ellis:1998}, SOAs have proven useful for various types of all-optical operation, including decoders logic\cite{Lei:2012, Cabezon:2014}. Amplified spontaneous emission (ASE) has, however, limited the cascadability of such SOA based all-optical switches. To combat these shortcomings, signal regeneration\cite{Leclerc:2005, Vivero:2009, Nguyen:2011} schemes were introduced, however increasing the power required to operate such devices. Further, the recovery time of the SOA limits its performance, but it has been shown that more than 320 Gbit/s\cite{Liu:2007} all-optical switching is possible, with some implementations enabling even the Tbit/s domain\cite{Ju:2005}. While SOAs are mostly based on III-V semiconductor technologies, such as Indium Phosphide (InP), efforts are being made to integrate them with silicon photonics, including with a focus on optical computing\cite{Tossoun:2023}. Apart from ASE, there are further downsides to using SOAs, such as high power consumption and heat output, as well as fabrication cost associated with the low volume production of III-V compared to CMOS in the current landscape. To address these issues for use in optical computing, alternative materials are explored, most notable 2D-materials\cite{Liu:2020}. The strong optical nonlinearities in 2D materials lend themselves well for all-optical switching\cite{Li:2014} and these all-optical switches can be integrated directly with the common Silicon-on-Insulator (SOI)\cite{Ooi:2017, Wang:2020} and Silicon Nitride (SiN)\cite{Demongodin:2019, Lukosius:2023} photonics platforms. The nonlinearities explored in graphene, such as saturable absorption, allow for extremely fast switching\cite{Alexander:2015,Demongodin:2019}, with recovery times well below $<1ps$, offering a path forward from SOAs.

The discussion on memory is twofold. On the one hand, we need an element that is able to store information to be read (and ideally written to) optically, on the other hand we require a decoding scheme to address the individual bits. For a comprehensive review on optical memory technologies, we refer to Lian etal. \cite{Lian:2022} and Alexoudi etal.\cite{Alexoudi:2020}. The topic of addressing memories is less well explored. While the previous all-optical switching devices do lend themselves to be used this way, it requires considerable power for large address spaces\cite{Cabezon:2014}. Instead, a more elegant approach is to use passive devices as optical decoders for addressing memory\cite{Vagionas:2013, Simos:2023} and considerable progress is being made using phase-change materials (PCM) for optical memory with novel addressing schemes\cite{Narayan:2022,Yang:2023,Sunny:2023}. For an overview of the use of PCMs in photonics now and in the future, we refer to the review by Prabhathan etal. \cite{Prabhathan:2023}.

For our discussion on all-optical computers, further elements apart from logic and memory are required. For synchronization purposes, a stable clocked laser signal is required, such as those based on microcombs\cite{Ulanov:2024}. Furthermore, we require various crossbars throughout our architecture, which are well explored in recent photonics literature\cite{Feldmann:2020, Xiao:2023, Zhu:2024}.

\section{All-Optical General-Purpose Computing}\label{sec:arch}

To fully understand our approach to all-optical computing, we use a top-down structure for the discussion and start on a computer architecture level. In the field, a bottom-up approach is the more common, but, as we will see, the architecture approach makes some technologies viable that would not necessarily be considered bottom-up. So far, in optical computing, the architecture considerations have barely taken into account actual usability and feasibility of an all-optical computer. This has led to the very valid criticism of the approach and garnered a host of misconceptions regarding the technology. As these architectural misconceptions are seldom addressed, we do so here, focusing on the following 3 that we hope to clear up throughout this paper:

\begin{enumerate}
	\item Misconception that any general purpose processor, like a modern GPGPU or Intel's x86 platform, requires billions of transistors to perform the operation out of necessity, and that this amount of transistors is needed to create competitive devices.
	\item Misconception that processor need to be 32- or 64-bit to handle modern day and future workloads.
	\item Misconception that a volatile random access memory (RAM or similar) is a strict requirement in any computer architecture and for high performance computing.
\end{enumerate}

\subsection{Efficient General-purpose without billions of transistors}\label{sec:subleq}

The first misconception we want to clear up, is the notion, that one needs billions of transistors to implement a feasible computer. This misconception stems from the false analogy that is often drawn from electronics. Electronics has the advantage, that it is easy to integrate a lot of devices on a small area. However, the easy availability of this resource to integrated circuit designers automatically leads to an excessive overuse. Much like the number of pages in a book or the number of lines of code neither reflect on the quality, nor functionality, neither does the count of transistors. Modern CISC (complex instruction set computer) architectures, such as x86, contain a multitude of instructions that are barely used, if at all, by most pieces of software, and were included to speed up very specific edge cases. The renewed interest in RISC (reduced instruction set computer) architectures is a clear indication, that this excess is not necessarily a good development.

Instead, we want to highlight the other end of the spectrum, namely how many devices it would take to implement a fully general-purpose processor using a minimal amount of devices. First, by general-purpose we mean a processor that can either run any type of software natively or emulated, given enough memory. In the extreme, that would include the ability of running an operating system with a 3D game loaded – i.e., the canonical question of “Will it run Doom\textsuperscript{\texttrademark}?”.

General-purpose computing merely reflects the fact, that the device is itself able to fully simulate a Turing machine (TM)\cite{Turing:1936}, given an infinite amount of memory. For obvious reasons, here we will refer by TM to a device that is practically Turing-complete (TC) but does not have an infinite amount of memory. In turn, the TM itself is the simplest computer able to run Doom\textsuperscript{\texttrademark}. While such a TM would be easily implemented using a handful of logic elements as an all-optical computer, it is highly impractical and not easily compatible with classical software. Instead, we investigate the simplest TC RISC architecture that we can implement, namely SUBLEQ. It should be noted, that SUBLEQ acts merely as a toy example for this demonstration and discussion, the final architecture will make use of a true RISC architecture.

SUBLEQ\cite{Mavaddat:1988, Mazonka:2011} is a one instruction set computer (OISC)\cite{Jones:1988} and simply performs a “subtract and branch if less-than or equal to zero” and we utilize a slightly modified version (with an additional $C$ parameter, which will become apparent in the memory section later): 

\begin{verbatim}
	Memory[C] = Memory[A] - Memory[B]
	If Memory[C] <= 0 Then Goto J
\end{verbatim}

or written short in the following notation:

\begin{verbatim}
	SUBLEQ [A], [B], [C], J
\end{verbatim}

where \texttt{[A]}, \texttt{[B]} are the memory locations of the operands of the subtraction, with the result stored in \texttt{[C]} ($C = A – B$), and \texttt{J} being the new jump location, if the new value of \texttt{[C]} is equal or less than zero. The word SUBLEQ can of course be omitted, as every instruction is the same, but kept here for readability. 

It should come to no surprise, that SUBLEQ can be turned into one of its sub-operations by calling it in the correct pattern. Furthermore, it can, of course, also perform operations that it doesn’t directly implement, like addition (subtraction by two’s complement), left shifting (adding a value binary onto itself), multiply (conditional left shifting) and any type of bitwise operation. We also allow SUBLEQ to dereference pointers using \texttt{[[X]]} to avoid the need for self-modifying code (Table \ref{tab:subleq})\footnote{Instructions are ordered in the AT\&T notation. \texttt{IP+1} refers to the next instruction location and not an actual addition, non-bracketed numbers refer to constants, bracketed numbers refer to constants referenced in memory, \texttt{[A*]} to a COW location, if COW is enabled (see Section \ref{sec:cow}), \texttt{[T]} to some temporary or ``dump`` location and \texttt{[SP]} to a pointer to the top of some stack memory. Also note, that the x86 \texttt{MOV} typically is not able to move from memory to memory, but only able to act as either a load (move from memory to register) or store operation (move from register to memory).}. This makes SUBLEQ one of the simplest instruction set architectures (ISA) to implement, while retaining the full ability to run, for example, RISC-V RV32I instructions with a simple translation layer\cite{Klemmer:2022}. 

\begin{table}
	\begin{center}
		\caption{Examples of the most common instructions emulated in SUBLEQ.}
		\label{tab:subleq}
		\begin{tabular}{ l | l }
			Instruction & SUBLEQ Equivalent \\
			\hline
			\scriptsize\texttt{MOV [A], [B]} & \scriptsize\texttt{SUBLEQ [A], [0], [B], IP+1}\\
			\hline
			\scriptsize\texttt{SUB [A], [B], [C]} & \scriptsize\texttt{SUBLEQ [A], [B], [C], IP+1}\\
			\hline
			\scriptsize\texttt{JMP [J]} & \scriptsize\texttt{SUBLEQ [0], [0], [T], J}\\
			\hline 
			\scriptsize\texttt{ADD [A], [B], [C]} & \scriptsize\texttt{SUBLEQ [0], [A], [T], IP+1} \\
			& \scriptsize\texttt{SUBLEQ [B], [T], [C], IP+1} \\
			\hline 
			\scriptsize\texttt{INC [A]} & \scriptsize\texttt{SUBLEQ [A], [-1], [A*], IP+1} \\
			\hline 
			\scriptsize\texttt{PUSH [A]} & \scriptsize\texttt{SUBLEQ [SP], [1], [SP*], IP+1} \\
			& \scriptsize\texttt{SUBLEQ [A], [0], [[SP*]*], IP+1} \\
			\hline 
		\end{tabular}
	\end{center}
\end{table}

As noted earlier, SUBLEQ or an ISA in general does not imply a von Neumann architecture, as \texttt{[A]}, \texttt{[B]}, \texttt{[C]} and \texttt{J} can all reside in different memory types and be accessed through various means and paths. How memory is accessed is defined by the physical realization of the processor and the compiler’s implementation of memory allocation patterns only. ISAs themselves are a feature of both von Neumann and post-von Neumann general-purpose architectures. They come in many shapes and forms and can be as simple as programmable routing configurations for a neuromorphic circuit or as complex as an x86.  For the sake of simplicity, we do not address those that are exclusively data driven and do not rely on an ISA at their core, even though they do pose very interesting candidates for an all-optical implementation, such as a cellular automaton. 

\begin{figure}[!t]
	\centering
	\includegraphics[width=3.4in]{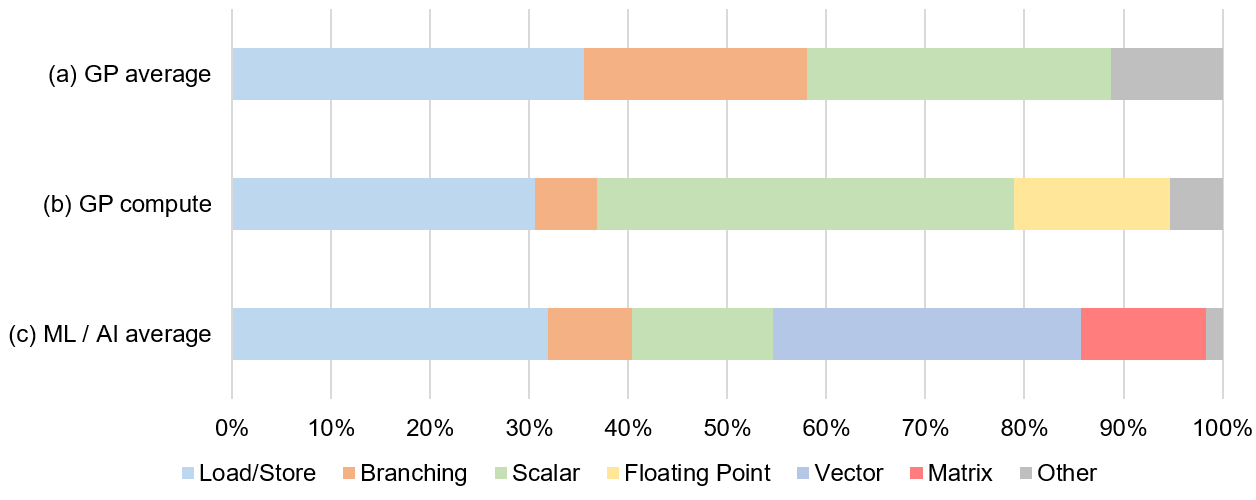}
	\caption{Instruction statistics in (a) average general-purpose (GP) processing on an x86 processor running a Windows operating system, executing day-to-day tasks\cite{Ibrahim:2010}, (b) average computing workloads, such as video decoding, sorting algorithms or simulations \cite{Kankowski:2009, Ndu:2012, SPEC} and (c) average AI / ML workloads, including training and inference\cite{Chen:2019}.}
	\label{fig:stat1}
\end{figure}

From Table \ref{tab:subleq} we see that many of the common instructions are implemented efficiently even in SUBLEQ, however not all. For any ISA it is important to understand the statistics of instructions used for a target application and the required cycles to execute such an instruction, in order to determine which to implement explicitly/natively for optimization. The common figure of merit is to calculate the overall processing time that would be lost based on instruction statistics and cycles required (CPI) or $t_{\text{ins}} = n_{\text{ins}} \cdot CPI_{\text{ins}}$. We break down the instruction statistics based on use-cases in Figure \ref{fig:stat1}. For general-purpose tasks on an x86, the most common instructions are \texttt{MOV}, \texttt{ADD}, \texttt{PUSH} and those discussed in Table \ref{tab:subleq}. But the availability of billions of transistors in electronic ICs has allowed the x86 instruction set to bloat to over 1000's of instructions optimizing even the rarest of use. So many in fact, that there exist dedicated fuzzing tools to identify hidden instructions in the space of approx. 100,000,000 possible ones\cite{Domas:2017}. This can be considered excessive, as the formula for $t_{\text{ins}}$ should hint at, and are primarily micro-optimizations. 

The reasonable lower bound for CPI of a serial instruction is 1. Apart from adding more processing cores, another measure to increase performance is to parallelize on an instruction level, such as by introducing single instruction multiple data (SIMD)\footnote{For example, a 32-bit value can also be vectorized as 4 individual 8-bit values and a SIMD instruction, such as a multiplication, would act on all 4 in parallel.}. But parallelization in general always faces diminishing returns, as the serial speed of the process eventually dictates the maximum speed-up that can be gained (Amdahl's Law\cite{Amdahl:1967}). The situation, however, becomes interesting when considering AI use-cases. Here, the diminishing returns are pushed further and further back by the need for more and more vectorization. In Figure \ref{fig:stat1}, we see that for AI the majority of operations are still centered around vector-matrix-multiplication (VMM) even on an instruction set with SIMD, as was the case here. For our architecture, it is obvious that to enable efficient AI tasks of the future, implementing VMM is a must, as both $n_\text{VMM}$ and $CPI_\text{VMM}$ will remain high. This guides our development of our optical computing architecture, in implementing instructions prioritized by the performance gain to be expected.

As the previous discussion showed, SUBLEQ is, of course, not the target realization for optical computing. Its purpose is to showcase the simplest form a general-purpose optical computer could take and an intermediary step we take. It can be implemented with less than 100 logic gates and, given enough memory, able to emulate a full x86 with a graphics card running Windows and Doom\textsuperscript{\texttrademark} loaded, while crunching AI models as a background task (admittedly all extremely slowly). It is meant to drive home one point, that every instruction or every logic gate added to a processor thereafter, has the sole purpose of optimization. Later, we explore the memory side of this point, but for the computation side of an optical processor, the trade-off is exclusively between performance and floorspace - and not ability.

\subsection{``2-Bit is not enough``?}

Marketing around processors often cited the increase in bits (for example 16-bit instead of 8-bit) as a very important improvement, until the almost unanimous adoption of 64-bit for desktop CPUs the past 10 years. With the ongoing AI-fueled race towards new chips, the bit width has again become a focus. But when referring to the bit width, it is often not exactly explained, which part of the processor is meant by $N$-bit. As an example, a 4-bit processor is quite definitely referring to the data words instead of the address bit width, as it would be quite a limited implementation to only be able to address 16 ($2^4$) different memory locations. Such is the case, for example, for the Intel 4004 with its 4-bit wide data words, yet 12-bit wide addresses. This misconception can lead to the impression, that a 2-bit processor is limited in its memory, but there is nothing preventing a 2-bit processor of having a 64-bit address space or more.

Another misconception regarding the bit width is the capability of the processor. A 2-bit, 4-bit or even 8-bit processor seems limited in its accuracy, as it seems to be only able to represent a very small range of numbers. However, even during the age of these processors in the 70’s, it was standard practice to represent wider bit numbers, such as 32-bit, using a combination of 2-, 4- or 8-bit numbers. In its simplest form, 4 unsigned 8-bit integers can represent a 32-bit unsigned integer. Of course, then arithmetic operations require more steps to perform a calculation involving two composed 32-bit numbers, but this now becomes merely a discussion of performance and not ability. Even today in the age of 64-bit number representations, the need for larger numbers above 64-bit is still common in scientific research. And as before, these big numbers are represented using concatenated memory locations to enable arbitrary-precision integers and decimals. The previous discussion on Turing machines should have also dismantled this misconception.

With the choice of bit width on data words being merely a choice on performance (and with floorspace being the trade-off), the question is, what is the ideal width? 64-bit is a great general-purpose candidate, but requires an almost $O(n^2)$ number of logic elements to implement even the simplest arithmetic function, making it less ideal for pure throughput. In deep learning models, the current push is towards AI quantization\cite{Gholami:2021}. This quantization process reduces the typical 32-bit floating point precision of deep learning models all the way down to 4-bit values\cite{Sun:2020, Dettmers:2023}, without a significant loss in precision of the model. Not only does this drastically reduce the size of the models, such as for LLaMA\cite{Dettmers2:2023} and many other large language models\cite{hugging}, but also improves the performance of the computation. The current sweet spot for AI quantization is converging to a 4-bit and 8-bit integer paradigm\cite{Baalen:2023}. 

The use of 4-bit and 8-bit integers in AI over the classical 32-bit floating point representations make a very strong case for optical computing, as it would otherwise be quite unfeasible to implement the full IEEE 754 floating-point standard\cite{IEEE:2019} in a PIC and software emulation too performance heavy. However, 4- and 8-bit integer operations can be implemented economically. For the few tasks (see also Figure \ref{fig:stat1}) that require real numbers, alternatives to floating-point, such as posits, are proving to be more accurate and easier to implement in hardware\cite{Gustafson:2017} or software emulation is valid option.

For an all-optical processor using the current generation of PIC technologies, a 16-bit wide arithmetic unit strikes a good balance between performance and gate count. 16-bit integers are wide enough, that the majority of the numbers that are currently represented by 32-bits or 64-bits in software experience a minor performance hit, as they rarely exceed a value of 65,535 (examples include most counters in loops, array indices, Boolean values or characters). Should the need arise to represent a 32-bit or 64-bit value, the performance loss is on a manageable order of magnitude for individual arithmetic operations. Furthermore, 16-bits would also enable SIMD of four 4-bit values or two 8-bit values, adding an additional layer of parallelization for AI discussed earlier.

As important as improving arithmetic throughput is, the most important aspect about our choice of 16-bit width is, that it is sufficient to enable fixed width instructions. This is vital to reduce the complexity, as it reduces one of the more problematic parts to implement on a processor, the instruction decoder, to a mere look-up table. Furthermore, it greatly simplifies the design of any potential instruction memory and addressing, instruction pointer arithmetic or instruction fetch operations. This balance between performance and gate count for the arithmetic unit and the induced simplicity of the peripheral units (instruction decoder, controller and counter) are what motivate us to declare 16-bit an ideal width for optical digital circuits to strive for at this point in time. 

\subsection{Computing while Propagating \& Reversible Computing}

Optical parametric nonlinearities allow for the creation of near instantaneous all-optical switching devices\cite{Boyd:2020}, granting optical computing the moniker ``time-of-flight computation``. This also means, that there is quasi no latency induced by the switching itself, unlike in electronics, where capacitance effects in the transistor do affect timing. These parametric nonlinearities, if used for computing, also allow the theoretical implementation of adiabatic and reversible logic. Again, in contrast to the capacitance and resistive effects faced in electronics. 

\begin{figure}[!t]
	\centering
	\includegraphics[width=3.4in]{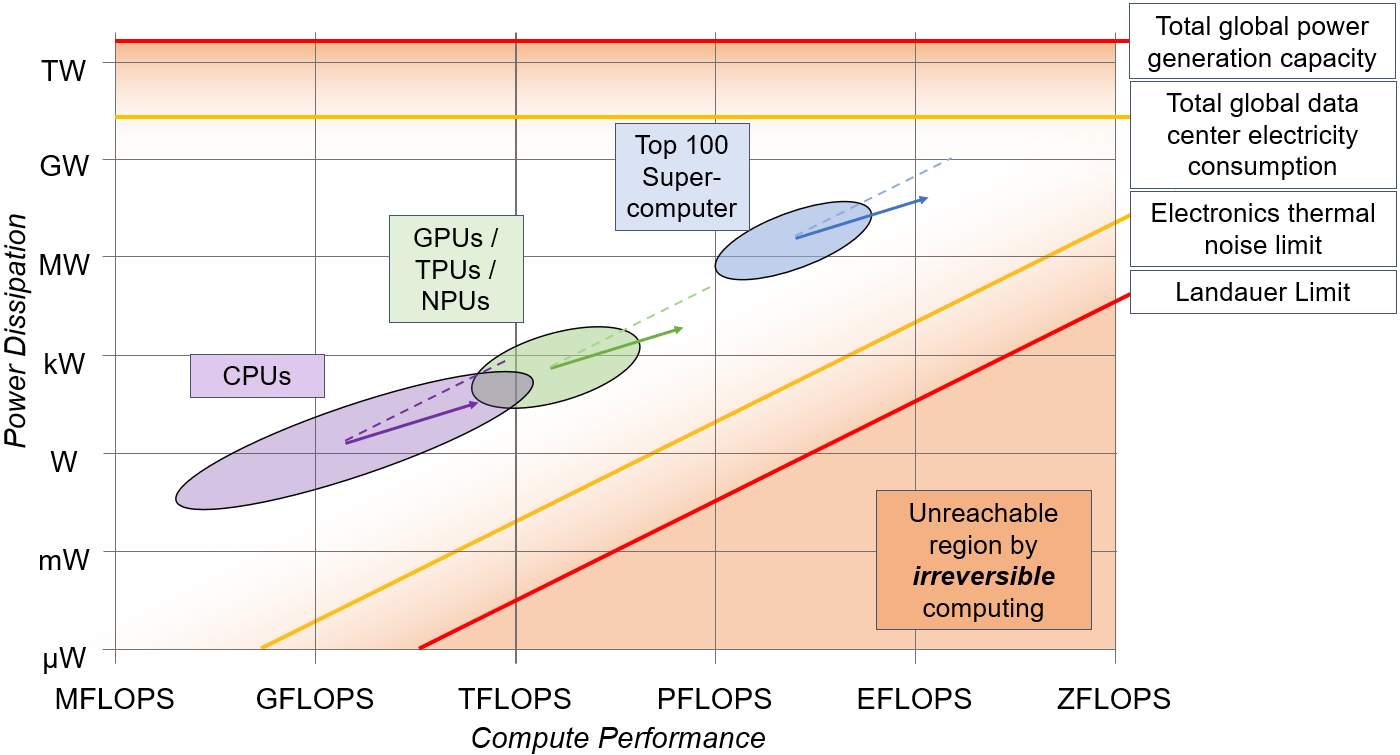}
	\caption{Visualization of Landauer's limit based on \cite{Frank:2020} and extended with IEA data from 2022\cite{IEA} and recent benchmarks. The floating-point operations per second (FLOPS) for GPUs, NPUs, TPUs and CPUs are based on the reported non-sparse tensor values. The angle of the diagonal orange, red and dotted lines represent constant FLOPS/W, with the arrows indicating that the overall trend in future devices to be more efficient (lowering FLOPS/W). The orange zones indicate no-go areas for the current irreversible approach to computing, but which can be crossed into with reversible computing. }
	\label{fig:reverse}
\end{figure}

This has major implications on the type of computation possible in the optical domain. Namely, optical computing is one of the few technologies that is in theory capable of reversible computing\cite{Lecerf:1963, Bennett:1973}, both in the physical reversibility, as well as the logical reversibility sense. This would allow going below Landauer's principle of $E \geq k_B T \ln 2$ energy consumed per irreversible bit operation, where $k_B$ is the Boltzmann constant. While computing in general is still far from this limit, optical computing does offer a unique path towards adiabatic computing. In Figure \ref{fig:reverse} the Landauer limit is visualized to highlight the merits for pursuing reversible computing as a whole.

A positive side-effect of reversible computing is also the lifting of one of Miller's criteria (see Table \ref{tab:miller}), namely the need for fan-out, which is implicitly forbidden in reversible computing. For the coming decade, however, we do not expect to have the required fabrication methods and optical materials to induce the previously mentioned parametric nonlinearities. While it will require considerable research and development over the next decade to enable reversible computing, it is important that we structure the new generation of computing architectures around this idea. Even before reaching full reversibility, the principles offer advantages, one of which, the idea of having no side-effects in the processor, will play a vital role in the coming section. 

\subsection{Write-Once-Read-Many is key}\label{sec:cow}

Write-once-read-many (WORM) memory allows, as the name implies, being written to once and being read as many times as possible. Optical examples include CDs, DVDs and Blu-Rays, as well as the currently developed 5D storage\cite{Zhang:2016, Wang:2021}. This non-volatile memory storage has many advantages, such as the ability to store information for decades without consuming power, allowing near instantaneous read-out, very fast writing, offering heightened data security and the ability to store vast amounts of data (100's terabyte in a small area\cite{Zhang:2016}). While the obvious downside is the single write operation, it should be noted, that these memories can usually be reset, but that this reset ability is on much longer timescales. A WORM memory that allows resetting the memory at a longer time scale ($>100\times$) than it takes to read or write to the first time, we will refer to as RWORM (resettable WORM).

The relevance to optical computing becomes clear when we once again consider the statistics in Figure \ref{fig:stat1}. While not explicitly stated, the ratio of load to store operations is actually very high, with most executed code using $\sim 7\times-20\times$ the amount of loads\cite{Ndu:2012}. This is amplified, for example, in AI inference, where the model usually doesn't change during execution, but needs to be read in its entirety, leading to load operations outweighing store operations significantly. This would make optical RWORM an attractive proposition. Combined with an all-optical addressing scheme with its near instant decoding, the gain in memory access bandwidth would be immense, as there is no limit on parallel read-out. Furthermore, the latency would only be limited by the physical distance to the compute unit requesting the data. I.e., an all-optical RWORM implementation can be expected to have a constant round-trip time of $\sim 1ns$ for a worst-case distance of the memory location to the processor of $\sim 70mm$ using SiN waveguides for transport. This is in stark contrast to modern electronic HBM and DDR-SDRAM memories, with read-out speeds of $73.3ns - 106.7ns$\cite{Huang:2020}, especially considering that most HBM sit right next to the processing unit.

With such fast read-outs and low latency, that leaves the question open for writing and resetting RWORMs. As discussed in \cite{Kissner:2023}, an RWORM memory with a copy-on-write (COW) mechanism is fully capable of being used in the same manner as a volatile storage, given enough memory and regular resets of ``used`` memory locations. It is interesting to note, that the reset ability is not even a required feature for general-purpose processing, as a write-once Turing machine is equivalent in its processing to a universal TM\cite{Wang:1957, Minsky:1961}, clearing up misconception 3. But much like SUBLEQ, this construction merely proves the ability and not the feasibility. To make RWORM a feasible storage for general-purpose computing, we identified the following:

\begin{itemize}
	\item A selective COW mechanism that only performs a copy-operation on the affected memory region.
	\item A software paradigm that minimizes the use of volatile storage.
\end{itemize}

For a selective COW mechanism, the implementation can be either in hardware (for example by adding an additional bit to each memory word that indicates it has been written) or a compiler solution that keeps track of the memory locations at compile-time. The change in hardware would obviously require more advanced logic to be implemented physically, which we want to avoid, so the preference for optical computing should be a compiler solution with a corresponding programming paradigm. Luckily, these COW mechanisms are well explored in literature, as they are prevalent in cache mechanisms, as well as operating system kernels\cite{Bovet:2005}.

As for a programming paradigm that is based on non-volatile memory, it is a perfect fit for a functional language, such as Haskell, Common Lisp, Erlang and F\#. But even imperative languages, such as C++, Python or Rust can be used in a functional style. The idea of using functional programming to inspire a new computer architecture was first proposed by John Backus in his 1977 Turing award lecture\cite{Backus:1978}. One important aspect of functional programming is, that all functions are deterministic (or pure) and have no side-effects. These principles from lambda calculus\cite{Barendregt:1984} lead to a programming style that disallows the use of mutable values, i.e. once a value has been assigned, it can’t be changed. This is an exact fit for the non-volatile memory explored here. There are two important things to note, however. First, even though the language itself forces immutable values, that does not necessarily mean, that the current compilers for electronic processor architectures implement it as such - mainly because they do not have the same restrictions as we do. For our purposes, this means we need specific compiler back-ends to ensure immutability even once translated to the instruction level. Secondly, even though the majority of a software can be written in a functional style, the I/O (network, sensors, ...) are not considered pure, and in a real production software, there are almost always un-pure parts. As a side note, the idea of pure functions and the functional programming scheme is well aligned with the previously discussed notion of reversible computing. Reversible computing, of course, must also be without side-effects and every function (or logic operation) completely deterministic. While reversible computing goes a step further by demanding bijectivity, lambda calculus is probably the closest classical paradigm we currently have.

Both these improvements on the RWORM principle are compiler based: a functional style programming and utilizing a selective COW tracking mechanism for un-pure parts. They make RWORM a feasible memory storage using familiar technologies, without relying on actual volatile memory. Of course, this comes at slight loss of performance, as functional programming is less explored and currently produces slower executing code than imperative languages. This performance hit, however, can be negated by adding in additional volatile memory and allowing certain parts of the code to be un-pure. 

As an example, consider a game like Doom\textsuperscript{\texttrademark}. While the overwhelming majority ($>98\%$) of the memory used would not need to change throughout execution (for example all the textures, level geometry or music), handling the character's and enemies' position, as well as health would ideally be implemented using mutable values. While it is certainly possible to do this functionally, it would essentially lead to one COW operation per frame per value in the worst case. Introducing a small amount of volatile memory would greatly improve performance. A similar case can be made for AI inference, which eclipses AI training by a an estimated factor of up to $1386:1$ even today\cite{Chien:2023}, where the model itself does not change, making it ideal for a RWORM memory and an all-optical implementation. 

\subsection{The World isn't Binary}

Optical computing is often misunderstood to be synonymous with either quantum computing or analog computing. And there is good reason, as interference of optical waves or single photons allow to us perform interesting arithmetic and quantum effects efficiently. So, it makes sense to harness these effects in the general-purpose context here as well, as there is no law stipulating that a digital processor must use exclusively digital operations. 

As an example, consider all-optical implementations of digital-to-analog (DAC)\cite{Kong:2017} and analog-to-digital (ADC)\cite{Patent:2009} converters. First, we convert two $N$-bit digital signals into analog all-optically. Next, we use those two analog signals and interfere them. Finally, we convert the analog signal back into digital all-optically using an optical ADC. We now essentially created a full $N$-Bit adder circuit. But there is a caveat: while a DAC can be implemented efficiently, an ADC is a complex device not just in optics, but even in electronics. The worst-case implementation of comparing each signal level individually implies a complexity of better than $O(2^N)$. With current optical nonlinearities and photonic integration, we can't reliably go above 4-bit for all-optical ADCs in the near future. Here, however, we are interested in the discussion, if adding analog compute to our architecture using all-optical DAC/ADCs has a benefit at all. The benefits possible are multitude: saving on floorspace by reduced number of logic gates, energy efficiency and improved latency. The simplest example of an analog $N$-bit adder to replace a digital adder, has minor benefit, as the implementation of the optical DAC and ADC itself requires almost as many non-linear optical elements as the equivalent optical logic. But there are many operations that do greatly benefit from an analog computing implementation in a DAC/ADC context, such as:

\begin{itemize}
	\item Vector Matrix Multiplications
	\item Fourier Transforms
	\item Non-Linear Activation Functions for Deep Learning
\end{itemize}

In a very similar pattern, optical continuous variable measurement-based quantum computing (CV-MBQC)\cite{Asavanant:2022} can benefit from this hybrid computing approach as well, by enabling an all-optical feed forward\cite{Sakaguchi:2023}. The current electro-optical approach has a few problems in regard to latency, as the feed forward operation, or more specifically, the non-linear arithmetic involved, needs to be performed electronically. This is an obvious mismatch between the time it takes to compute this function and the delay the optical modes need, making CV-MBQC scaling using feed-forward a technical challenge. But this can be avoided by doing both optically using all-optical phase sensitive detection\cite{Takanashi:2020}.

Our goal of this section was to highlight the fact, that general-purpose optical computing can both greatly benefit from results in analog optical computing, as well as contribute to quantum optical computing in profound ways. An additional benefit is, that all 3 domains, analog, digital and quantum in optical computing are implemented using very similar PIC technologies, such as SiN waveguides. This is a unique feature throughout the field of computing, be it electronics or some other medium. In some sense, the all-optical control flow mechanisms enabled by digital optical logic circuits act as a familiar janitor to the analog and quantum approach and enable a truly cross-domain approach to optical computing. 

\subsection{Architecture beyond von Neumann}\label{sec:XPU}

\begin{figure}[!t]
	\centering
	\includegraphics[width=3.2in]{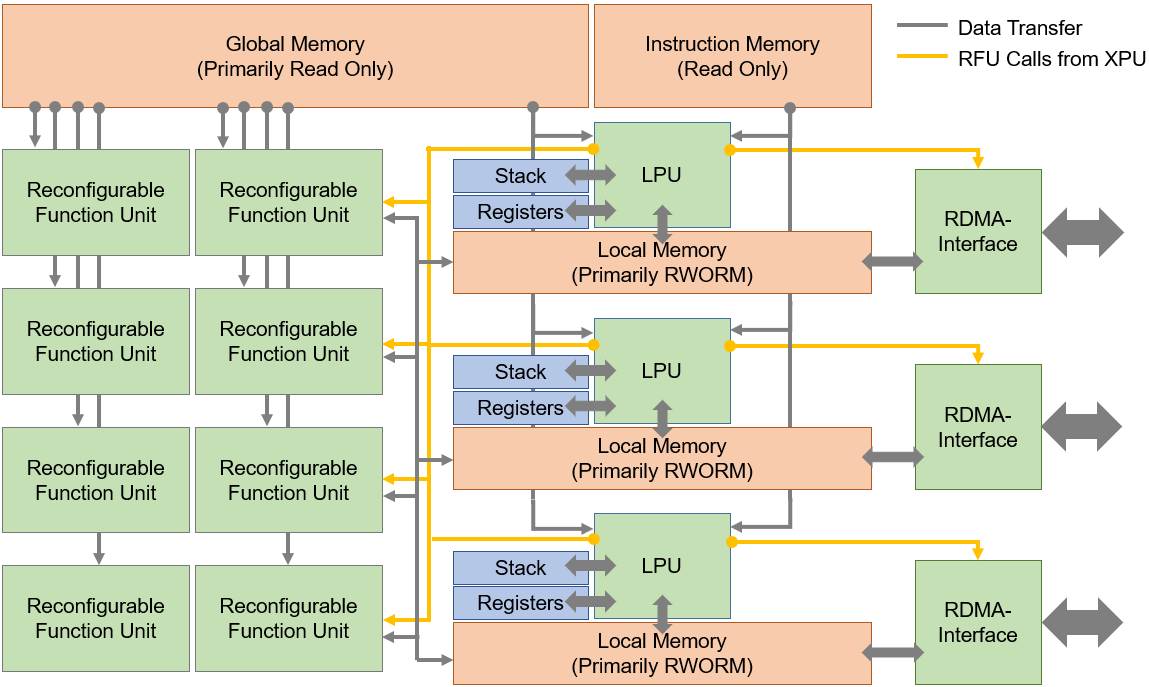}
	\caption{All blocks in this architecture, as well as all data paths (data transfer and RFU calls) are all-optical. Orange highlights memory storage primarily based on non-volatile technologies (RWORM, ROM) and blue volatile memories. The green blocks indicate compute units and the XPU can communicate with the outside world using interfaces, such as the RDMA interface depicted here.}
	\label{fig:arch}
\end{figure}

In this section, we want to combine the previously explored architectural considerations into one coherent platform. Inspired by the previous section, we will refer to this platform as our optical cross-domain processing unit or XPU. Figure \ref{fig:arch} highlights the overall organization. 

The core of the XPU is the purely digital feed-forward logic processing unit (LPU), which contain no memory of their own and implement the typical features found in a general-purpose processor: instruction decoder, arithmetic logic unit (ALU), instruction pointer logic and memory controllers. All the typical registers one would find in a CPU are decoupled and external in our LPU. Multiple of these digital LPUs can populate the XPU. Each of these 16-bit units implement a reduced instruction set based on a load/store architecture and simple integer arithmetic, with one additional specialized instruction called reconfigurable functional unit call (\texttt{RFU}). In its simplest form, the SUBLEQ processor used as a toy example here could be used as an LPU, if extended by an \texttt{RFU} instruction and the ability to access all the different memory types. But in actual implementations of this architecture, more potent RISC ISAs will be implemented for the LPU.

The \texttt{RFU} instruction allows each LPU to access systolic array-inspired RFU-blocks, which are configured to fit the target use-case. The RFUs act as processing blocks in the functional programming sense and are equivalent to the functional forms $f$ introduced by Backus\cite{Backus:1978}. As we intend to optimize the XPU for use in AI applications, the majority of the RFUs will be VMM accelerators. From an implementation standpoint, it matters not, if these RFUs are digital optical circuits or, as is the case for the VMMs, analog optical VMMs with all-optical DAC and ADC. The RFUs neither receive nor send their input and outputs directly to the LPU and instead communicate using LPU-defined memory locations. This allows tightly coupling and daisy chaining the execution of the RFUs without the need to involve the LPU, apart from triggering the execution and collecting the results. It is equivalent to function composition $f\circ g$ in the functional style\cite{Backus:1978}. 

There are multiple memory types throughout the XPU. Following the ideas of a Harvard architecture, the instruction memory is physically decoupled from any data and read-only, which greatly improves access time, efficiency, as well as security. Each LPU has an attached register bank for storing intermediate results, as well as a stack memory for short term use. Both these storages are implemented using volatile memory technologies and reduced to a bare minimum. In addition, the LPUs have their own local memory, which is either fully RWORM or a hybrid RWORM-volatile storage. The purpose of this memory is to exchange information between LPUs, between LPUs and RFUs and between RFUs. Finally, we have a large ROM-RWORM hybrid storage referred to as our global memory. Primarily meant as a read-only memory for large amounts of data (for example an AI model for inference), it can be written to if absolutely needed. This multitude of memories and the parallel nature of the memory access available in optics essentially nullifies the von Neumann bottleneck\cite{Backus:1978}.

To enable communication beyond the XPU, interfaces can be added, such as a remote direct memory access (RDMA) network controller mimicking a subset of the InfiniBand\cite{TA:2023} standard\footnote{Akhetonics is part of the InfiniBand Trade Association and intends to push for a minimal subset of the specification aimed at all-optical implementations.}. The Ethernet and IP standard is too bloated for an all-optical integration at this point in time and would require either too much effort to realize in hardware or sacrifice too many clock cycles for a software centric implementation. Furthermore, InfiniBand and RDMA are a great fit for our architecture, as functionally, these interfaces can be implemented as any other RFU and addressed by the LPUs as such, simplifying the integration. As a matter of fact, we enforce that every interface shall be implemented as an RFU.

\begin{figure}[!t]
	\centering
	\includegraphics[width=3.2in]{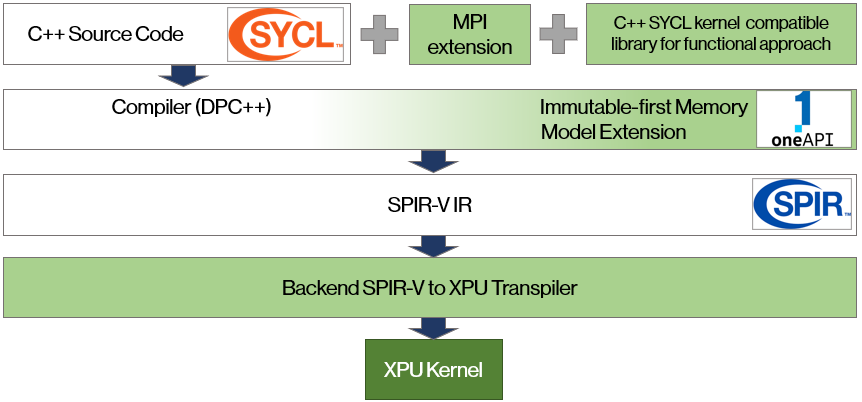}
	\caption{Compiler stack to enable optical computing using XPUs in a HPC environment. The green highlights indicate software and libraries that must be specifically implemented for the memory allocation patterns from Section \ref{sec:cow}, but are mostly abstracted away from the programmer to allow a seamless adoption.}
	\label{fig:compiler}
\end{figure}

On a software level, we ensure the XPU is compatible with existing code-bases and abstract away the finer details of the memory allocators and the functional approach highlighted in Section \ref{sec:cow} through the compiler. It will handle most of the complications surrounding RWORM memory, as well as handling the multitude of memory locations, but we intend to continue promoting a more functional mindset while programming. To illustrate this, the full compiler stack is shown in Figure \ref{fig:compiler}. 

In this stack, we follow the HPC cluster use-case with a central host machine (not necessarily optical) and multiple systems implementing the optical XPU architecture presented here, attached as accelerators through RDMA connections. A developer would write their HPC code using the SYCL\cite{SYCL} standard in C++. They would define the boilerplate code to run on the host machine and define individual SYCL kernels to be run on the XPUs, as well as the messaging fabric using a library based on MPI\cite{MPI}. For optimization purposes, both the MPI library, as well as any supporting libraries for the SYCL kernel need to adhere to memory allocation principles presented. The compiler (for example DPC++\cite{OneAPI}) would take this SYCL C++ code and compile the program for the host machine, as well as SPIR-V\cite{SPIRV} kernels for each XPU. Our custom compiler backend ingests the SPIR-V code and produces the optimized executable kernel for our XPU. In all, it should be clear, that both the architecture and the compiler go hand in hand when defining a new processor, especially when doing so beyond the classical von Neumann approach.

Finally, we address the topic of cycles. In a classical electronic CPU, an instruction is usually executed over multiple cycles. This can be due to the unavailability of data in the cache, requiring a fetch from memory or simply because certain instructions are more complex arithmetically.

In our all-optical computing architecture, we introduce the mantra of \textit{everything happens in one cycle}. This means, that the entire pipeline of fetching and decoding the instruction, fetching memory, performing arithmetic, logic and RFU operations, as well as the final write-back\footnote{We do not expect the written data to be readable in one cycle.}, all happen in one cycle deterministically. With time-of-flight computation, this is greatly simplifies things, as it removes the need to store a state for indetermined amounts of time and the need for logic circuit to handle the complex await mechanisms. With near instantaneous switching speeds, the total time for each cycle is solely dependent on the length of all the interconnects and can be determined from the physical design of the all-optical computer. The total time for a cycle is a sum of the longest waveguide path ($l_\text{max}$) for memory round-trips, LPU execution and RFU calls and $t_{\text{cycle}} = \frac{l_\text{max} n}{c}$. For SiN waveguides, this would result in a path length of approx. $15cm$ for a $1ns$ cycle time. And in our all-optical processor, our mantra guarantees all memory read operations to return after 1 cycle or $1ns$ in this case. This is in stark contrast to electronics, where memory access can induce $>200$ cycles and $>100ns$ in latency even on a high end processor\cite{Wong:2010}.

The real beauty of time-of-flight computing and the reversible approach is, that without side-effects we can introduce very dense pipelining of threads. Here, the recovery time of the switching element dictates how dense we can pack threads into the pipeline (see also Figure \ref{fig:delay}). For a nonlinearity based on 2D materials, for example, with recovery times $\ll 1ps$\cite{Xing:2010}, these threads can be spaced at $1ps$. With $1ns$ fixed cycle time and up to $1,000$ pipelined threads, it allows a single core XPU to execute at 1 tera operations per second (TOPS) in non-vectorized mode and ignoring RFUs. When including vectorization and RFUs for VMM for example, the TOPS equivalent scales proportionally. Furthermore, there is no limit on the bandwidth of memory read operations in an optical setting, allowing true tera bit per second (Tbps) transfers.

\section{Photonic Implementation}\label{sec:impl}

The previous architecture discussion allows us to make informed decisions on which photonic elements are viable to implement our processor and which are not. Our goal here is to show a real physical implementation of these principles. While we won’t be implementing the full, 16-bit ISA with all the features previously discussed, we will be implementing a 2-bit variant of SUBLEQ for demonstration purposes, which can be used as a fully functional all-optical CPU. We begin by introducing our all-optical logic, followed by the memory and the processor. It should be noted, our purpose here is to demonstrate the feasibility of optical digital computing on a high level. 

\subsection{All-Optical Logic}

We presented a variety of all-optical decoding schemes in Section \ref{sec:sota}. To test the feasibility of our approach, we implement a hybrid decoder using a mixture of the discussed XGM approach using SOAs in a monolithic InP platform and passive SiN elements for interconnects and tuning elements, which both individually have been shown to be compact and fulfill a majority of the criteria laid out by Miller in similar applications. Our construction largely follows the canonical signal-probe MZI setup\cite{Pan:1995} in a monolithic InP platform\cite{Tekin:2000, Soares:2019}. To reduce the contributions of ASE, saturable absorbers in the same platform are introduced in a 2R configuration\cite{Vivero:2009}, leading to a trade-off between total power consumed by the SOAs and the resulting extinction ratio (ER). While the 2R setup enables ER improvements of $>4 db$ for input signals with an ER $<4 db$, achieving an output extinction ratio throughout the system of $>5 db$ is sufficient for our purposes. 

\begin{figure}[!t]
	\centering
	\includegraphics[width=3.4in]{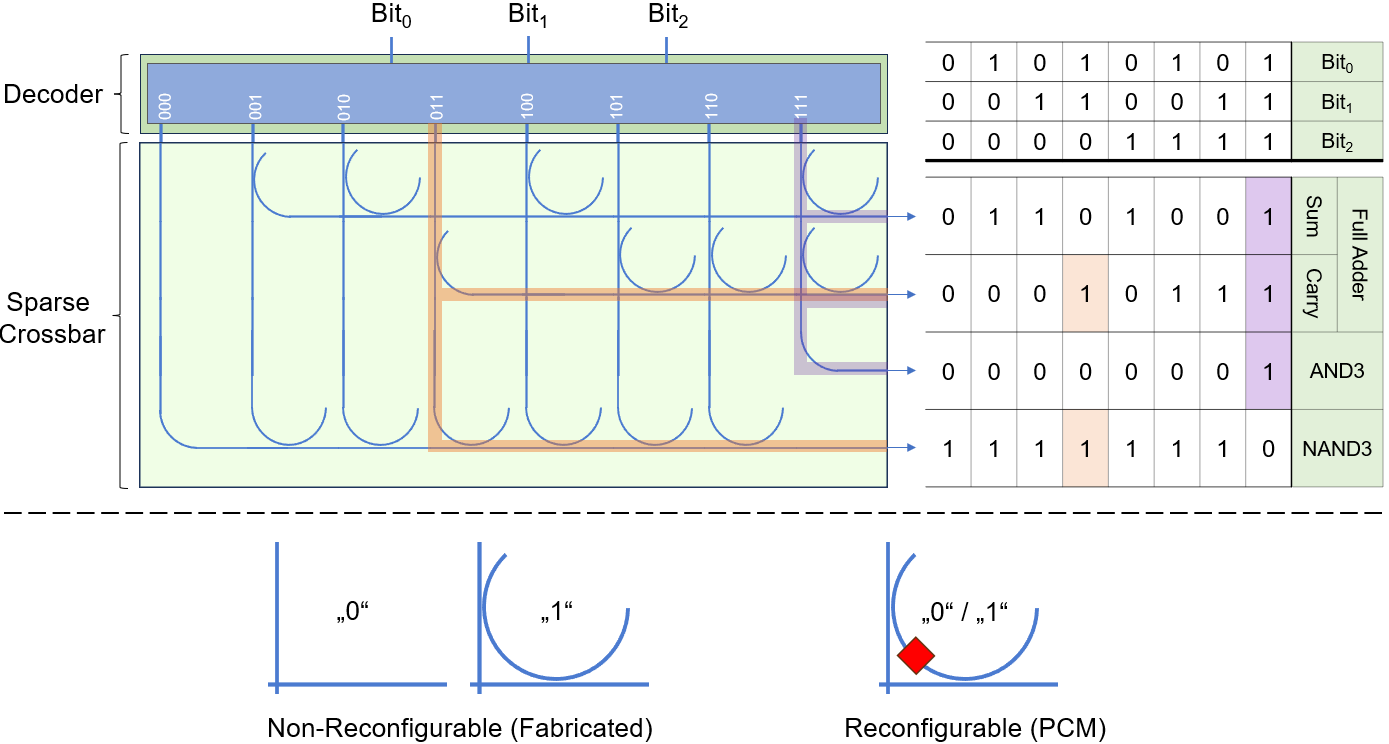}
	\caption{Working principle of the sparse crossbar in conjunction with an all-optical 3x8 decoder for digital logic. Here demonstrated 3 use-cases, as a full-adder with 2 outputs, a 3-input \texttt{AND} and a 3-input \texttt{NAND} gate. The orange and purple beams highlight the path of the individual look-ups. Also included is a view on the 2 different possibilities of implementing the crossbar, either as non-reconfigurable (permanently set at fabrication) or as reconfigurable (using a naive PCM implementation).}
	\label{fig:crossbarlogic}
\end{figure}

The resulting one-hot\footnote{Alternative to binary encoding, where one bit is set to $1$ and all others to $0$. The position of the bit defines the value. For example \texttt{101} (5) in binary is equivalent to \texttt{00100000} in one-hot.} decoded digital signal allows us to perform arbitrary logic functions using a crossbar as shown in Figure \ref{fig:crossbarlogic}. Usually, these crossbar configurations are used for analog processing, however by utilizing it in binary, the structure simplifies considerably. Essentially, the crossbar becomes a 1-to-1 reflection of the binary truth table, as follows,
\begin{equation}
	\begin{split}
		f_\text{dec} ([\text{Bit}_0, \text{Bit}_1, \text{Bit}_2]) &= \vec{x}_\text{1-hot} \\
		f_\text{xbar} (f_\text{dec} ([\text{Bit}_0, \text{Bit}_1, \text{Bit}_2])) &= \textbf{M}_\text{xbar} \cdot \vec{x}_\text{1-hot}
	\end{split}\label{eq:crosslogic}
\end{equation}
where, when we use the orange example of Figure \ref{fig:crossbarlogic}, with the 3 input bits being \textbf{\texttt{0}}, \textbf{\texttt{0}}, \textbf{\texttt{1}}, i.e. $f_\text{dec} (\textbf{\texttt{011}}) = [0, 0, 0, 1, 0, 0, 0, 0]$, we get
\begin{equation}
	\left[
	\begin{array}{cccccccc}
		0 & 1 & 1 & 0 & 1 & 0 & 0 & 1  \\
		0 & 0 & 0 & 1 & 0 & 1 & 1 & 1  \\
		0 & 0 & 0 & 0 & 0 & 0 & 0 & 1  \\
		1 & 1 & 1 & 1 & 1 & 1 & 1 & 0 
	\end{array}
	\right] \left[
	\begin{array}{c}
		0 \\
		0 \\
		0 \\
		1 \\
		0 \\
		0 \\
		0 \\
		0 \\
	\end{array}
	\right] = \left[
	\begin{array}{c}
		0 \\
		1 \\
		0 \\
		1 \\
	\end{array}
	\right]
\end{equation}

The multitude of $0$ entries in the matrix allows for much more compact designs, as the logical $0$ values do not need to be physically implemented (i.e., there is no crossbar at that intersection). For the automated design pipeline (Figure \ref{fig:atet}), this is an important consideration, as in an all-optical setting, the canonical \texttt{NAND} gate is the most inefficient and to be avoided during logic synthesis, whereas an \texttt{AND} gate or direct implementation of a full-adder is very efficient.

A further advantage of this structure of all-optical logic is, that multiple results can be generated in parallel, without the need for additional decoders or dedicated logic gates. This, for example, allows us to generate the output of a full-adder, which would require a multitude of logic gates, using a single decoder and crossbar, while still enabling us to output additional logic functions using that same input. Furthermore, by making the matrix $\textbf{M}_\text{xbar}$ in Equation \ref{eq:crosslogic} tunable, for example through the use of PCM, a logic function can be reconfigured after fabrication if needed, much like an FPGA. Since the crossbar is essentially digital (only allowing $0$/off or $1$/on values) and is only changed occasionally, if ever (comparable to burning an EEPROM for an update), it acts robustly against small errors in fabrication or material variation. A compact and robust, yet highly inefficient design is shown in Figure \ref{fig:crossbarlogic}. Essentially each Decoder $f_\texttt{dec}$ (performing 1-hot encoding) coupled to a crossbar $f_\texttt{xbar}$ (linearly transforming the 1-hot encoding to the desired output) acts as a look-up table (LUT) or read-only memory (ROM). 

Another optimization to consider is the use of each crossbar using multiple wavelengths simultaneously (as hinted at by the purple and orange in Figure \ref{fig:crossbarlogic}). Since we are computing in binary and in a one-hot encoding, there is no need for exact control of interference pattern and merely ``light or no light``, simplifying the design. But operating using multiple wavelengths would require a compatible decoder $f_\texttt{dec}$ scheme, which is highly non-trivial. 

\subsection{Optical Memory}\label{sec:optmemimpl}

For this demonstration, we require 3 different memory types: Read-only memory, RWORM memory and register memory. The stack and global memory of Figure \ref{fig:arch} are not required.

The implementation of the ROM is fairly straightforward, as the previous crossbar can be used as a LUT. For registers, the simplest implementation is the delay line, but ideally having reamplification, reshaping and retiming (3R). Retiming is partially performed by the clocked logic itself, thus we focus on 2R. This we realize using an in-line 2R configuration as presented in \cite{Vivero:2009}, which, while not energy efficient, can be readily implemented in a PIC. An alternative for the registers would be using a more robust flip-flop configuration\cite{Pleros:2009}. 

\begin{figure}[!t]
	\centering
	\includegraphics[width=3.2in]{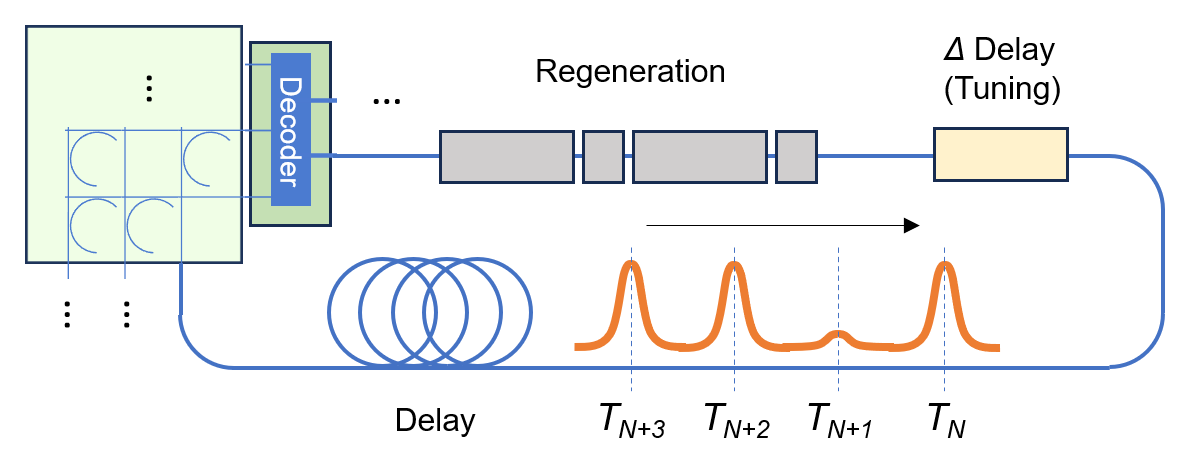}
	\caption{Delay line based register. The length of the delay is equal to one compute / clock cycle, allowing for multiple binary signals ($T_N, \cdots, T_{N+3}$ in this example) to reside in the waveguide, each storing a bit of data for a specific thread $T_i$ and enabling multithreading. A logic circuit decides the new value of the register bit each cycle. The spacing of the individual thread timings is proportional to the reset speed of the signal regenerator and logic - here shown as a 2R as in \cite{Vivero:2009}. For synchronization purposes, a $\Delta \text{Delay}$ can be tuned.}
	\label{fig:delay}
\end{figure}

Delay lines, however, do offer a unique feature by not holding a state, in that they can store a chain of signals in its entire length. This allows for the realization of the pipelining protocoll discussed for our XPU (see Figure \ref{fig:delay}). Even at a $1 ns$ cycle time, we are able to run up to 40 threads in parallel through time multiplexing using the InP-based regenerator \cite{Liu:2007, Vivero:2009}, without the need to add additional physical registers or other structures. 

Finally, we have the PCM-based RWORM memory. To understand our choice, we want to highlight a historical curiosity. DVD-RAM\cite{dvd:2005} from 1996, not to be confused with a regular DVD, DVD-R, DVD-RW or other variants, was a unique storage medium meant to compete with HDD and flash memory. It was able to perform over 100,000 re-writes by using the PCM GST, had a very high storage density and beat its competing platforms (HDD and flash) on most metrics, most notably cost per GB. Using zoned constant linear velocity (ZCLV)\cite{dvd:2005} operation and its faster seek, DVD-RAM was usable, as the name suggest, for random access, allowing it to be used like a modern USB-drive. But even with its superior technology, this was also its downfall, as it made it incompatible with most regular DVD/DVD-RW drives. 

While the DVD-RAM technology sadly failed due to low adoption and with development discontinued in the early 2000's, it still serves our discussion as a usable RWORM drive. Of course, 28 years later, one would not opt for a mechanical drive for addressing, but it was an optically readable and writable medium, making the material platform perfectly suitable for an all-optical implementation. Ignoring the mechanical seeking aspects, read-out of each memory bit was instant (reflecting a beam off the disk). But the latency of writing a bit into GST using a laser is, by optical computing standards, slow ($>20 ns$\cite{Faneca:2020}). For our purposes, thus, it can be viewed as a RWORM memory with the ability to reset regions on a longer time scale. An actual modern implementation would, of course, use a higher density medium, include multiple read-write heads, all-optical addressing and remove any mechanical aspects to go beyond the mere 22.16 Mbit/s speeds of the early 2000's. 

This being said, we opted to incorporate the GST PCM directly on SiN\cite{Faneca:2020}, which was performed by the University of Southampton in conjunction with CORNERSTONE\cite{Littlejohns:2020}. This allows us to test the principle on a small scale and use it directly for our LPU / CPU. A new alternative is to use low-loss PCMs with large index contrast, such as $\text{Sb}_2\text{Se}_3$\cite{Delaney:2020}, to increase efficiency. The PCM-based device can be used both as a flashable ROM written to before computation starts, as well as the previously discussed RWORM. The actual writing of the PCM all-optically in this demonstration is hinted at and more information on the all-optical write scheme will be published separately.

\subsection{The CPU}

A single fully monolithic PIC for a general-purpose, all-optical processor is not feasible at the current time. Instead, we use heterogeneous integration methods, such as chip-to-chip coupling using grating couplers and fibers or butt coupling. Furthermore, this has the advantage of simplifying debuggability of the individual components and increases overall yield. The processor was designed using our custom digital photonics circuit design pipeline. Inspired by the electronic-equivalent OpenLane and OpenROAD\cite{OpenLane, OpenROAD} projects, the pipeline follows an overall similar process and re-uses existing tools, such as Yosys for register-transfer-language (RTL) synthesis. However, it diverges significantly in the physical design process, as outlined in \cite{Kissner:2023}.

\begin{figure}[!t]
	\centering
	\includegraphics[width=3.4in]{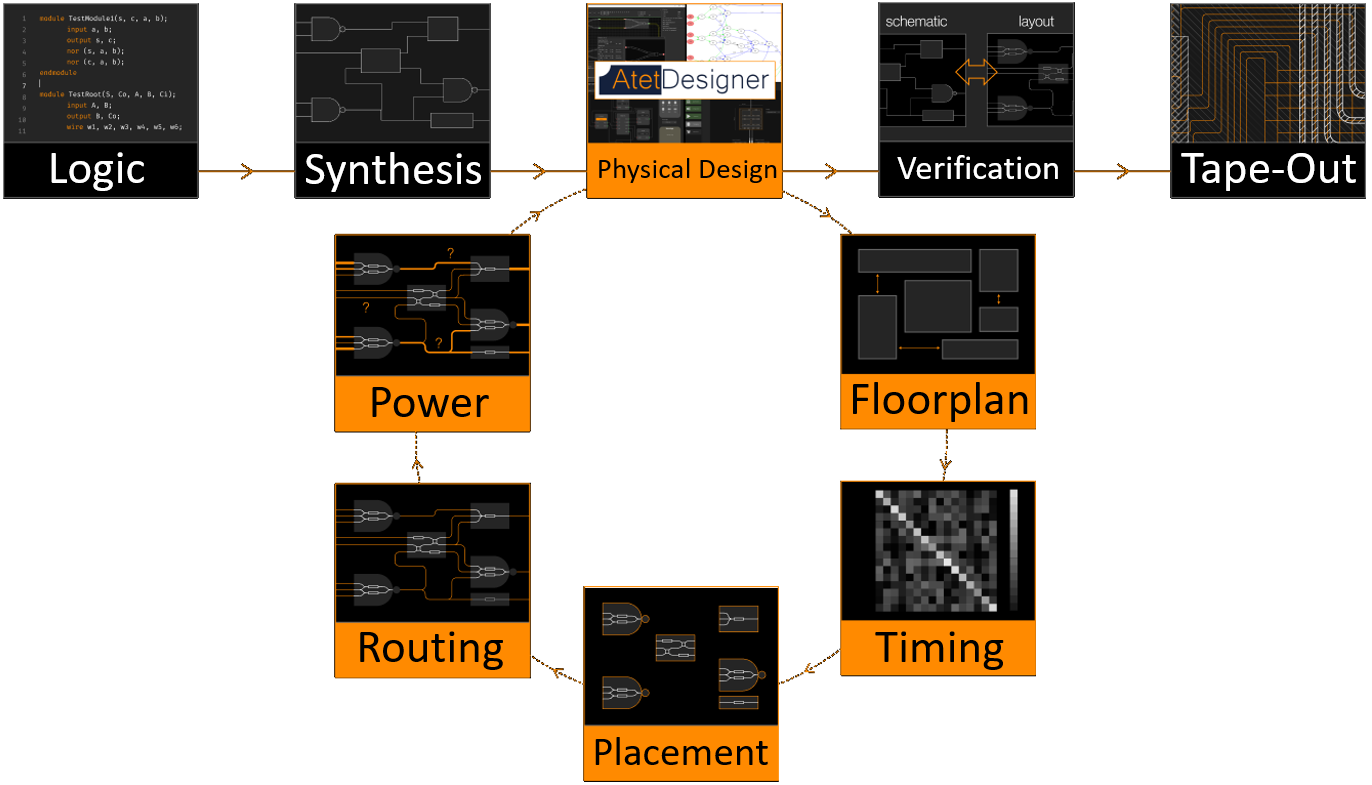}
	\caption{The fully automated RTL-to-GDSII flow, with our custom physical design process for all-optical digital logic highlighted in orange, making it the first all-optical very large scale integration (VLSI) pipeline for optical digital computing. The existing open-source tools, such as Yosys\cite{Yosys} and the OpenROAD project\cite{OpenROAD,OpenLane} remain usable in an all-optical setting apart from the highlighted physical design stage.}
	\label{fig:atet}
\end{figure}

To enable modularity, on a hardware level, our current iteration is implemented in a similar form factor to TTL-based processors in the 60’s and 70’s\cite{Lancaster:1974}, such as the 7400 series by Texas Instruments. These TTL ICs typically only contained a few logic gates or memory cells and were interconnected to form a full processor. In much the same spirit, we limit the amount of crossbar logic per PIC and interconnect the full processor using a multitude of these devices, according to the results of Figure \ref{fig:atet}. This allows easy modularity and the possibility to perform bit slicing (i.e., constructing a 4-bit device out of multiple 2-bit devices). While the constant in-and-out coupling to the PIC introduces avoidable losses, it greatly increases our ability to troubleshoot. Furthermore, the purpose here is to show the ability for all-optical general-purpose compute and not a highly optimized implementation. The ability to troubleshoot and tune the logic circuits is further enhanced, by deploying programmable interconnects\cite{Bogaerts:2020}, such as our iPronics SmartLight system\cite{iPronics}, which greatly speeds up interconnecting the devices and tuning the needed delay lines.

\begin{figure*}[!t]
	\centering
	\includegraphics[width=\textwidth]{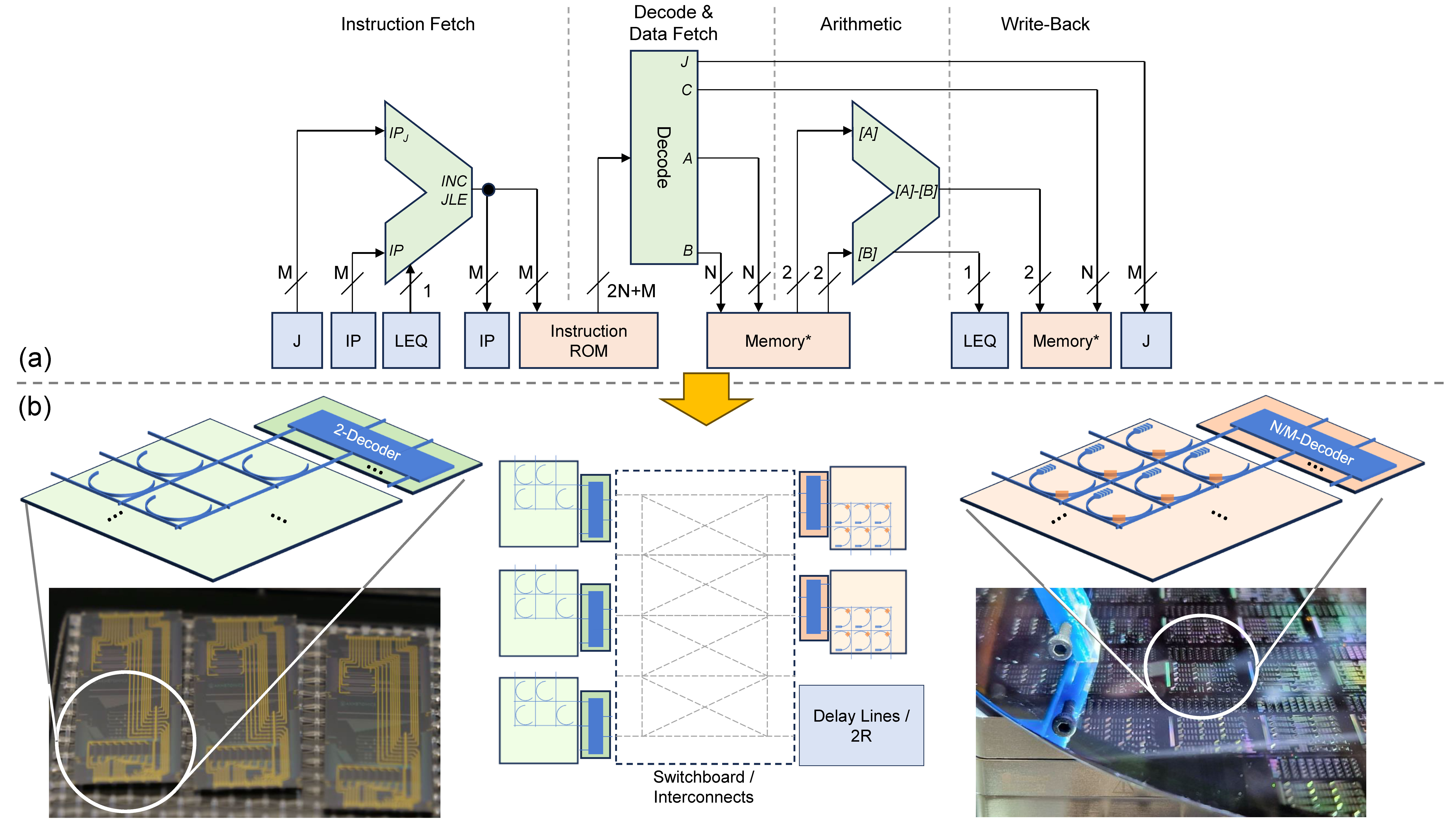}
	\caption{Implementation of the first all-optical CPU. (a) Shows the steps the SUBLEQ-based processor needs to perform, where green highlights the logic operations, orange the memory access and blue the registers. Here, $M$ denotes the width of instruction ROM address space, $N$ the width of the memory address space and the $*$ indicates, that these memories are not needed to be uniform and can be of any type (RWORM, volatile, etc.). (b) Shows the photonic implementation of (a) using the same color scheme. Highlighted here also the physical PICs, left using a SiN platform for the fixed crossbar and heaters for fine tuning and right the PCM-enhanced SiN for the memory, both of which are connected to a separate non-linear element for decoding. To enable flexibility and debugging, the devices are interconnected using discrete fiber optics in this demonstration.}
	\label{fig:ankh2}
\end{figure*}

On a logic level, our CPU implements the 2-bit SUBLEQ scheme outlined in \ref{sec:subleq}. The physical implementation of which is shown in Figure \ref{fig:ankh2}.

\subsection{Results}

\begin{figure}[!t]
	\centering
	\includegraphics[width=3.2in]{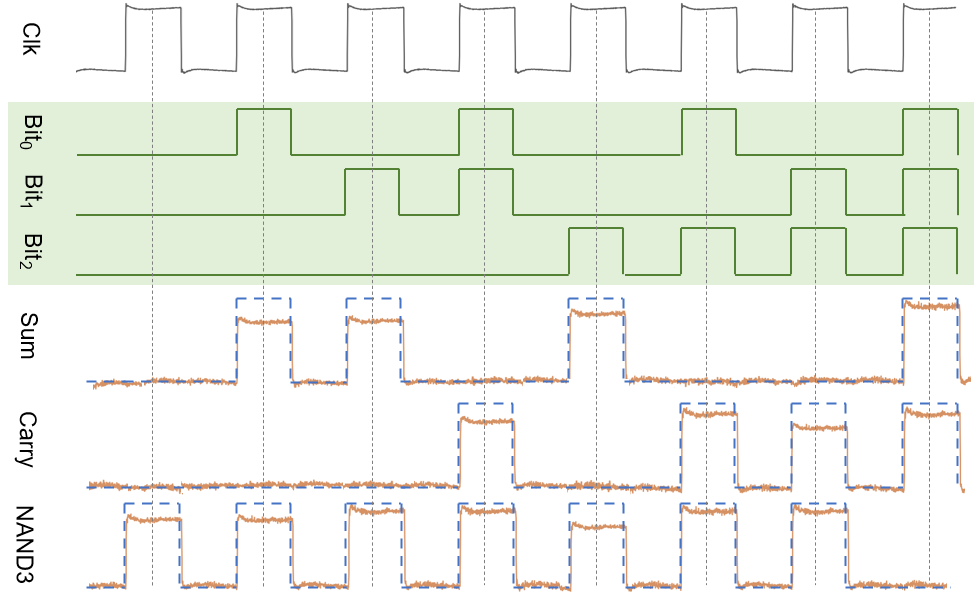}
	\caption{Response of the full-adder and \texttt{NAND3} of the sparse crossbar logic from Figure \ref{fig:crossbarlogic} prior to regeneration. The bit inputs and dashed lines are the expected waveforms, while the orange line indicates the actual response.}
	\label{fig:results}
\end{figure}

All tests were conducted on bare PIC dies to allow quick debugging. To allow for testing of multiple devices per PIC, the PCM-based RWORM chips use grating couplers and are expected to have higher losses in this iteration, than one optimized for insertion loss (see also Section \ref{sec:impl} in regards to design choices for the crossbar). The configuration of the test setup can be seen in Figure \ref{fig:ankh2}, where we note this is an abstract view and multiple decoders and crossbars were fitted per InP and SiN chip respectively.

Each PIC contains multiple copies of the used structures, coupled individually with grating couplers for characterization. The structures for the SiN PICs are: Waveguides, bends, PCM covered waveguides, grating couplers, variable directional couplers, edge couplers, terminators, crossings, heaters. The structures for the InP PICs are: Waveguides, bends, edge couplers, 1x2 MMIs, SOAs. Apart from the PCM covered waveguides, all components were designed according to the respective foundry's PDK and characterization of the PICs has concluded that sufficiently many operate within the given parameters of the PDK\footnote{Due to non-disclosure agreements with the foundries, exact results can not be published, but access to the PDKs can be obtained directly from the foundries.}. As indicated in Section \ref{sec:optmemimpl}, the results on the custom PCM-based components will be reported seperately, keeping the focus on the architecture here.

We begin by testing the compute functionality in CPU operation of the chip and plot the logic response each clock cycle in Figure \ref{fig:results}. Due to limitations in our test equipment, we used a sub-GHz clock for testing. Shown are the elements also depicted in Figure \ref{fig:crossbarlogic}, where we included the \texttt{NAND3} for generality. The varying height and time offsets of the logical \texttt{HIGH} signals is due to  the directional couplers used and the resulting coupling ratios. These can be fixed by introducing tuning elements, which we omitted for the entirety of the memory PIC and only partially included in the logic PIC.

Considering that a typical CMOS full-adder comprises of $28T$ (transistors), the depicted crossbar configuration could be said to be equivalent to $\approx 40T_e$ in electronics, factoring in the additional outputs. Our goal here is not to highlight the efficiency of all-optical logic implementations, but rather to point out the contrary: there is no value in comparing the abundant complexity of electronic circuit design with all-optical computing, as they follow completely different paradigms. A 1-to-1 comparison of integration density is not useful and undermines almost all of the advantages optical computing offers. For example, the $\approx 40 T_e$ estimate does not include additional logic outputs in the crossbar, use of wavelength multiplexed mode or the ability to multithread reusing the same circuitry. These strategies are hard to quantify in transistor equivalent, as they do not scale linearly if realized in an electronic circuit. That being said, for our case, each logic unit was designed for 8 usable logic outputs, 1 wavelength and 40 parallel threads, and we will use a conservative $\approx 1500 T_e$ equivalent for our discussion here.

The area size of the depicted logic circuit is split into 3 parts. For the 3-to-8 decoder, the area required in SiN was $9.88 mm^2$ and in InP was $14.9 mm^2$. Each crossbar cell in SiN had an area requirement of $0.12 mm^2$ (or $7.68 mm^2$ for all $8 \times 8$ cells). These numbers include all waveguides and interconnects. The structures were designed to be testable, with grating couplers at $250 \mu m$ pitch added throughout as debug ports and each crossbar cell having its own. Ignoring the test structures and optimizing for floorspace the total area is $3.7 mm^2$ for InP and $4.2 mm^2$ for SiN. 

While we do not promote the use of transistor equivalent as a metric, we do want to give some historic context on how our pure demonstrator CPU here stacks up to an electronic processor. The famed 6502 processor, one of the highlights of the 80's/90's that was used in anything from the Commodore 64 to a Nintendo (NES), had a transistor density of $194\;T/mm^2$. In the optimized form, our processor in its current iteration is able to reach a very similar density at $190\;T_e/mm^2$, but our processor shown here can do so running at a clock speed $1,000\times$ faster (1 GHz vs. 1 MHz) and $40\times$ multithreaded. Pushing the limits of SOA design\cite{Ju:2005}, this could even increase by another $25\times$. Again, the PPA metric\footnote{Power, Performance, Area(, Cost)} in semiconductors is always an apples to oranges comparison, especially in this case, but does provide perspective.

\section{Conclusion and Outlook}

We presented our all-optical, cross-domain XPU computing architecture for high-performance computing tasks. As part of the architecture, we demonstrated, for the first time ever, an implementation of a fully general-purpose CPU capable of operating all-optically. Unlike other optical computing approaches, ours requires no conversion of data to the electronic domain at any point and can run all-optically throughout each compute cycle. We showed how we realized this physical implementation and the individual building blocks (logic, memory).

An important part of the discussion was to clear up misconceptions surrounding optical computing, which we hope to have done so here. Our goal was and is to shift the discussion away from the often-questioned ability of optical computing to a discussion on performance. In this aspect, the goal was also to show the potential that optical computing has in regard to performance and efficiency, and that it is more than just a viable candidate for the future of computing. With a clear path towards trillions of operations and data transfers per second per core, reversible computing and cross-domain (combining digital, analog and quantum elements in one) computing, optical processors is positioned well to solve a lot of issues currently plaguing electronic compute.

We showed how the physical CPU presented here can be expanded on to be used as the logic processor (LPU) for the XPU computing architecture and gave a roadmap to realize a full XPU with a true RISC ISA instead of SUBLEQ. Our efforts are now aimed towards the next iteration of the decoder and realizing it using 2D materials for a massively more energy efficient and faster design compared to SOAs. We are also focused on expanding the decoding and logic scheme to much wider ranges. This all will allow us to reach ultra-high bandwidth computing, ultra-low latency memory running at minimal power in the near future.


\bibliographystyle{IEEEtran}
\bibliography{main}

\vfill

\end{document}